\documentclass[11pt]{article}
\usepackage[utf8]{inputenc}
\usepackage[left=1in,right=1in]{geometry}
\usepackage{authblk}
\usepackage[inline]{showlabels}
\usepackage[pdfstartview=FitH,pdfpagemode=None]{hyperref}
\usepackage{amsmath}
\usepackage{amsfonts}
\usepackage{youngtab}
\usepackage{cite}
\title{Note on generating functions and connected correlators of \\ 1/2-BPS Wilson loops in $\mathcal{N}=4$ SYM theory}
\author[1]{Anthonny F. Canazas Garay}
\affil[1]{Instituto de F\'isica, Pontificia Universidad Cat\'olica de Chile \authorcr Casilla 306, Santiago, Chile}
\author[2]{Alberto Faraggi}
\affil[2]{Departamento de Ciencias F\'isicas, Facultad de Ciencias Exactas, \authorcr Universidad Andr\'es Bello \authorcr Sazi\'e 2212, Piso 7, Santiago, Chile}
\author[3,4]{Wolfgang M\"uck}
\affil[3]{Dipartimento di Fisica ``Ettore Pancini", Universit\`a degli Studi di Napoli ``Federico II" \authorcr Via Cintia, 80126 Napoli, Italy}
\affil[4]{Istituto Nazionale di Fisica Nucleare, Sezione di Napoli \authorcr Via Cintia, 80126 Napoli, Italy}
\date{}
\numberwithin{equation}{section}

\newcommand{\rmd}{\,\mathrm{d}}

\newcommand{\Tr}{\operatorname{tr}}

\newcommand{\e}[1]{\operatorname{e}^{#1}}

\newcommand{\Laguerre}{\operatorname{L}}

\newcommand{\BesselI}[1][0]{\operatorname{I}_{#1}}

\newcommand{\vev}[1]{\left\langle #1 \right\rangle}

\newcommand{\Order}{\mathcal{O}}

\newcommand{\cWL}{\mathcal{W}}
\newcommand{\scWL}{\widetilde{\mathcal{W}}}

\newcommand{\kset}{\vec{k}}

\newcommand{\SG}{\mathfrak{S}}

\begin{document}
\maketitle
\begin{abstract}
The generating functions for the Wilson loops in the symmetric and antisymmetric representations of the gauge group $U(N)$ are expressed in terms of the connected correlators of multiply-wound Wilson loops, using ingredients from the representation theory of the symmetric group. This provides a proof of a recent observation by Okuyama.
As a by-product, we present a new calculation of the connected 2-point correlator of multiply-wound Wilson loops at leading order in $1/N$. 
\end{abstract}
\section{Introduction}\label{sec: intro}
Over the past decade, the study of non-planar ($1/N$) and quantum ($1/\sqrt{\lambda}$) corrections in AdS/CFT has become feasable thanks to the advent of localization techniques \cite{Pestun:2007rz}. These have provided a large number of exact results in gauge theories that lend themselves to asymptotic expansions, opening up the possibility of exploring the gravitational side of the correspondence beyond the leading order in which the classical action dominates.

In this context, the $\frac{1}{2}$-BPS circular Wilson loops in $\mathcal{N}=4$ SYM, in various representations of the $U(N)$ or $SU(N)$ gauge group, have played a central role \cite{Maldacena:1998im,Rey:1998ik,Erickson:2000af,Drukker:2000rr,Akemann:2001st,Hartnoll:2006is,Yamaguchi:2006tq,Gomis:2006sb,Okuyama:2006jc,Yamaguchi:2006te,Lunin:2006xr}. While the solution for the fundamental representation \cite{Drukker:2000rr} naturally furnishes an expansion in $1/N$, which is exact in $\lambda$, going beyond the leading order for arbitrary representations has proved to be more difficult, even though an exact formal solution is also known \cite{Fiol:2013hna}.
A systematic $1/N$ expansion for the $k$-antisymmetric representation $\mathcal{A}_k$ was obtained independently by Gordon \cite{Gordon:2017dvy} and Okuyama \cite{Okuyama:2018aij} using the method of resolvents \cite{Ambjorn:1992gw} in the Gaussian matrix model. The case of the $k$-symmetric representation $\mathcal{S}_k$ was addressed in \cite{Chen-Lin:2016kkk}.
The first $1/N$ correction for $\mathcal{A}_k$ was extracted in \cite{CanazasGaray:2018cpk} from the exact solution \cite{Fiol:2013hna}, but it is unclear whether a similar calculation is possible for higher orders in $1/N$. 
Efforts to compute $1/\sqrt{\lambda}$ corrections as well as 1-loop effective actions on the gravitational side of the duality include \cite{Forste:1999qn,Drukker:2000ep,Kruczenski:2008zk,Faraggi:2011bb,Faraggi:2011ge,Faraggi:2014tna,Forini:2015bgo,Faraggi:2016ekd,Horikoshi:2016hds,Forini:2017whz,Aguilera-Damia:2018bam,Aguilera-Damia:2018twq}.

After so much work dedicated to the circular Wilson loops, it is surprising that some general results can still be found.
Recently, Okuyama \cite{Okuyama:2018aij} observed that the $1/N$ expansions of the generating functions of Wilson loops in the $k$-symmetric and $k$-antisymmetric representations of $U(N)$ are related by\footnote{The definitions for the generating functions will be given in Sec.~\ref{sec: WL}.}
\begin{equation}\label{I:conjecture}
	J_S\left(z,\lambda,\frac{1}{N}\right)=-J_A\left(-z,\lambda,-\frac{1}{N}\right)\,,
\end{equation}
at least for the first few orders in the small-$\lambda$ series. He conjectured that this relation is exact in $\lambda$. In \cite{Fiol:2018yuc}, it was shown that this is actually an example of a general relation between the Wilson loops in representations corresponding to transpose Young diagrams, in this case,
\begin{equation*}
	\mathcal{A}_k=\left.\substack{\yng(1,1)\vspace{-0.15cm}\\\vdots\vspace{0.05cm}\\\yng(1)}\hspace{-0.05cm}\right\}\hspace{-0.05cm}k\leq N
	\qquad\textrm{and}\qquad
	\mathcal{S}_k=\underbrace{\Yvcentermath1\yng(2)\cdots\yng(1)}_{\displaystyle k}\,.
\end{equation*}
The proof in \cite{Fiol:2018yuc} was based on an expansion in terms of color invariants and is formally perturbative in powers of $g_{YM}^2$, or, equivalently, $\lambda$. An expansion in $1/N$ requires further work. Yet, the proof hints at a deeper, group-theoretical origin of the identity. 

Our work in this note is motivated by relation \eqref{I:conjecture}. We would like to use the representation theory of the symmetric group $\SG_n$ in order to express the generating functions $J_S$ and $J_A$ in terms of the connected correlators of multiply-wound Wilson loops in the fundamental representation of $U(N)$, which were (re)introduced by Okuyama in \cite{Okuyama:2018aij}. His observation \eqref{I:conjecture} then follows quite easily. It is interesting to note that the connected correlators are the natural objects to consider in the resolvent method, where they are obtained as a genus expansion \cite{Ambjorn:1992gw}. The use of (unconnected) correlators of multiply-wound Wilson loops has been advocated in work on 2-dimensional QCD \cite{Gross:1993yt} and $SU(N)$ Wilson loops \cite{Gross:1998gk}. From them, arbitrary representations can be obtained by linear relations. 
The exact matrix model solution for arbitrary representations \cite{Fiol:2013hna}, however, does not make direct use of these correlators, but employs the linear relation between the Schur basis of symmetric polynomials for arbitrary representations and the  monomial basis, for which the matrix model solution is fairly easy to compute. 

The present paper is organized as follows. We begin in Sec.~\ref{sec: WL} with a brief summary of the generating functions and connected correlators of circular Wilson loops. Sec.~\ref{sec: proof} is then devoted to the expansions of the generating functions for Wilson loops in the totally symmetric and antisymmetric representations of $U(N)$ in terms of the connected correlators. We conclude with a discussion of our results and possible future work. For completeness, we have collected in appendix \ref{app: comb} a few basic notions of combinatorial analysis, namely, permutations and partitions, which are necessary in the bulk of the paper. Also, we include in appendix \ref{app: W2} a short derivation of the connected two-point correlator based on the techniques developed in an earlier paper \cite{CanazasGaray:2018cpk}.
\section{Wilson loop correlators and generating functions}\label{sec: WL}
As shown in \cite{Erickson:2000af,Drukker:2000rr,Pestun:2007rz}, the $\frac{1}{2}$-BPS circular Wilson loop in $\mathcal{N}=4$ SYM theory with gauge group $U(N)$ can be computed exactly as the expectation value in a Gaussian matrix model, 
\begin{equation}
\label{WL:WL}
	\vev{\Tr_\mathcal{R} U} = \vev{\Tr_\mathcal{R} \e{\sqrt{\frac{\lambda}{2N}} M}}_{\text{mm}}~.
\end{equation}
On the left hand side, $U$ denotes the Maldacena-Wilson loop operator
\begin{equation}
\label{WL:MWL}
	U = P \exp \oint_C \rmd s \left( A_\mu \dot{x}^\mu + i \Phi |\dot{x}| \right)~,
\end{equation}
where $\Phi$ is one of the six scalars of $\mathcal{N}=4$ SYM, $C$ is a circular contour in space, $\mathcal{R}$ denotes a representation of $U(N)$, and $\lambda$ is the 't~Hooft coupling. On the right hand side of \eqref{WL:WL}, the matrix model expectation value is defined as 
\begin{equation}
\label{WL:MM}
	\vev{\mathcal{O}}_{\text{mm}} = \frac{ \int \rmd M \e{-\Tr M^2} \mathcal{O}}{\int \rmd M \e{-\Tr M^2}}~,
\end{equation}
with the integration taken over all hermitian matrices $M$.\footnote{In the case of $SU(N)$ the matrices must also be traceless.}

Instead of dealing directly with the individual Wilson loops, it is useful to consider generating functions from which these objects can be derived. For the case of the totally symmetric and antisymmetric representations of $U(N)$, the respective generating functions are defined as \cite{Hartnoll:2006is}
\begin{align}
	\label{eq: JS def}
	J_S\left(z,\lambda,\frac1N\right)&\equiv \frac1N \log \sum\limits_{k=0}^\infty z^k \vev{\Tr_{\mathcal{S}_k} U}=\frac{1}{N}\log\vev{\det\left(\sum_{k=0}^{\infty}z^k\e{k\sqrt{\frac{\lambda}{2N}}M}\right)}_{\textrm{mm}}\,,
	\\
	\label{eq: JA def}
	J_A\left(z,\lambda,\frac1N\right)&\equiv \frac1N \log \sum\limits_{k=0}^\infty z^k \vev{\Tr_{\mathcal{A}_k} U}=\frac{1}{N}\log\vev{\det\left(1+z\e{\sqrt{\frac{\lambda}{2N}}M}\right)}_{\textrm{mm}}\,.
\end{align}
Using the techniques of \cite{Fiol:2013hna}, the matrix model integrals can be computed exactly, yielding
\begin{equation}
\label{WL:J.MM}
	J_S = \frac1N \Tr \log \sum\limits_{k=0}^\infty z^k A(k)~,
	\qquad
	J_A= \frac1N \Tr \log \left[ 1+ z  A(1) \right]~.
\end{equation}
Here, $A(k)$ is the $N\times N$ symmetric matrix\footnote{$A(k)$ is related to the matrix $I(x,y)$ defined in \cite{CanazasGaray:2018cpk} by $A(k) = \e{\frac{k^2g^2}{2}} I(kg,kg)$, where $g^2= \frac{\lambda}{4N}$ is the matrix model coupling constant.}
\begin{equation}
\label{WL:A:expl}
	A(k)_{m,n} = A(k)_{n,m}= \sqrt{\frac{n!}{m!}} \e{\frac{k^2\lambda}{8N}} \left(\frac{k^2\lambda}{4N}\right)^{\frac{m-n}2}
		\Laguerre^{(m-n)}_n\left(-\frac{k^2\lambda}{4N}\right)~,
\end{equation}
where $\Laguerre^{(m-n)}_n(z)$ denotes the associated Lagurerre polynomial, and $m,n=0,\ldots, N-1$. 
In what follows, we adopt the simplified notation
\begin{equation}
\label{WL:A.notation}
	A_k = A(k)~.
\end{equation}

Besides the Wilson loop itself, another interesting object one seeks to compute is the correlator of $h$ multiply-wound Wilson loops in the fundamental representation of $U(N)$ (with identical circle contours $C$),
\begin{equation}
\label{WL:WLcorr}
	\vev{ \prod\limits_{i=1}^h \Tr U^{k_i}} = \vev{\prod\limits_{i=1}^h \Tr \e{k_i \sqrt{\frac{\lambda}{2N}} M}}_{\text{mm}}~.
\end{equation}
In \cite{Okuyama:2018aij}, Okuyama introduced the corresponding generating function\footnote{Deviating from Okuyama's notation, we have added the $h$-dimensional vector $\kset=(k_1,k_2,\ldots,k_h)$, which encodes the winding numbers of the loops, as a subscript to $G$. By definition, the Wilson loop correlators and the corresponding generating function are even under any permutation of the $k$'s.}
\begin{equation}
\label{WL:G.MM}
	G_{\kset} = \vev{\prod\limits_{i=1}^h \det \left(1 +y_i \e{k_i \sqrt{\frac{\lambda}{2N}} M} \right) }_{\text{mm}}~,
\end{equation}
such that
\begin{equation}
\label{WL:WLcorr.G}
	\vev{ \prod\limits_{i=1}^h \Tr U^{k_i}} = \oint \prod\limits_{i=1}^h \frac{\rmd y_i}{2\pi i y_i^2} G_{\kset}~.
\end{equation}
Again, an exact expression for the generating function is available, namely,
\begin{equation}
\label{WL:G.MM.sol}
	G_{\kset} = \det \left[ \sum\limits_{m=0}^h \sum\limits_{i_1<\cdots<i_m} y_{i_1} \cdots y_{i_m} 
	A(k_{i_1}+\cdots + k_{i_m}) \right]~.
\end{equation}

Finally, and more importantly for our purposes, Okuyama \cite{Okuyama:2018aij} also introduced the \emph{connected} $h$-point  correlator of multiply-wound Wilson loops, defined directly in the matrix model as
\begin{equation}
\label{WL:Gconn.def}
	\cWL_{\kset} = \vev{ \prod\limits_{i=1}^h \Tr U^{k_i}}_{\text{conn}} 
	= \oint \prod\limits_{i=1}^h \frac{\rmd y_i}{2\pi i y_i^2} \log G_{\kset}~.
\end{equation}
Using the exact form of the generating function \eqref{WL:G.MM.sol} one can compute \eqref{WL:Gconn.def} in terms of traces of products of the matrix $A_k$. The first few examples are \cite{Okuyama:2018aij}
\begin{align}
\label{WL:W1}
	\mathcal{W}_{(k_1)}&=\Tr\left[A_k\right]\,,
	\\
	\mathcal{W}_{(k_1,k_2)}&=\Tr\left[A_{k_1+k_2}\right]-\Tr\left[A_{k_1}A_{k_2}\right]\,,
	\\
\label{WL:W3}
	\mathcal{W}_{(k_1,k_2,k_3)}&=\Tr\left[A_{k_1+k_2+k_3}\right]-\Tr\left[A_{k_1+k_2}A_{k_3}\right]-\Tr\left[A_{k_3+k_1}A_{k_2}\right]-\Tr\left[A_{k_2+k_3}A_{k_1}\right]
	\\ \notag
	&+\Tr\left[A_{k_1}A_{k_2}A_{k_3}\right]+\Tr\left[A_{k_3}A_{k_2}A_{k_1}\right]\,.
\end{align}
A key observation is that these connected correlators have a systematic expansion in $1/N^2$ of the form \cite{Ambjorn:1992gw,Okuyama:2018aij} 
\begin{equation}
\label{WL:W.series}
	\cWL_{\kset} = N^{2-h} \sum\limits_{g=0}^\infty N^{-2g} C_{g,h}(\kset)~,
\end{equation} 
where $C_{g,h}$ is the genus-$g$ contribution.
\section{Generating functions in terms of connected correlators}\label{sec: proof}
Equations \eqref{WL:W1}--\eqref{WL:W3}, and their generalizations for $h>3$, express the connected Wilson loop correlators in terms of the traces of symmetrized products of the matrices $A_k$. Notice that each of these traces is of leading order $N$, whereas, according to \eqref{WL:W.series}, the $h$-point correlator is of order $N^{2-h}$. This means that as $h$ grows an increasing number of cancellations between the leading terms must occur. Therefore, it is more useful to consider the connected correlators as a basis and calculate the traces of symmetrized products of $A_k$'s in terms of them. The result of this exercise, up to $n=4$, is
\begin{align}
\label{P:W1}
	\Tr [A_k] &= \scWL_{(k)}~,\\
	\Tr [A_{(k_1}A_{k_2)}] &= \scWL_{(k_1+k_2)} - \scWL_{(k_1,k_2)}~,\\
\label{P:W3}
	\Tr [A_{(k_1}A_{k_2}A_{k_3)}] &= \frac12 \left[ 2 \scWL_{(k_1+k_2+k_3)} -3 \scWL_{(k_1+k_2,k_3)} + \scWL_{(k_1,k_2,k_3)} \right]~,\\
\label{P:W4}
	\Tr [A_{(k_1}A_{k_2}A_{k_3}A_{k_4)}] &= \frac16 \left[ 6 \scWL_{(k_1+k_2+k_3+k_4)} -8 \scWL_{(k_1+k_2+k_3,k_4)} 
	-3 \scWL_{(k_1+k_2,k_3+k_4)} \right. \\
	\notag &\quad \left. + 6 \scWL_{(k_1+k_2,k_3,k_4)} - \scWL_{(k_1,k_2,k_3,k_4)} \right]~.
\end{align}
%
On the left hand side, the parentheses represent symmetrization over the $k$'s. By the same token, $\scWL$ stands for the symmetrized version of the connected correlators $\cWL$. In either case, symmetrization is achieved using permutations $\sigma\in\SG_n$ acting on the index $i$ of $k_i$. For example, the simplest non-trivial case is
\begin{align}
\label{P:symm.ex}
	\scWL_{(k_1+k_2,k_3)} &= \frac16 \left( \cWL_{(k_1+k_2,k_3)} +\cWL_{(k_2+k_3,k_1)}+ \cWL_{(k_3+k_1,k_2)} \right.\\
	\notag &\quad \left.
	+ \cWL_{(k_1+k_3,k_2)} + \cWL_{(k_2+k_1,k_3)} + \cWL_{(k_3+k_2,k_1)} \right)~.
\end{align}

We now need to generalize formulas \eqref{P:W1}--\eqref{P:W4} to arbitrary $n$. To this purpuse, it is useful to introduce an additional piece of notation. Given an $n$-dimensional vector $\vec{k}=(k_1,k_2,\ldots,k_n)$ and a partition $\nu=(\nu_1,\nu_2,\ldots,\nu_c)$ of $n$ into $c$ parts, denoted by $\nu\vdash n|c$ (see appendix \ref{app: comb} for definitions and notation), we define the $c$-dimensional vector $\vec{k}_{\nu}$ by
\begin{equation}
\label{C:k.mu.def}
	\vec{k}_{\nu} = \left(\sum_{i=1}^{\nu_1}k_i, \sum_{i=\nu_1+1}^{\nu_1+\nu_2}k_i, \ldots,\sum_{i=\nu_1+\nu_2+\cdots+\nu_{c-1}+1}^{n}k_i\right)~.
\end{equation}
For example, for $\vec{k}=(k_1,k_2,k_3)$ we have
\begin{equation}
	\vec{k}_{(3)}=\left(k_1+k_2+k_3\right)~,
	\qquad
	\vec{k}_{(2,1)}=\left(k_1+k_2,k_3\right)~,
	\qquad
	\vec{k}_{(1,1,1)}=\left(k_1,k_2,k_3\right)~,
\end{equation}
corresponding to the three possible partitions of $n=3$. Notice the following special cases 
\begin{equation}
\label{C:k.mu.spec}
	\kset_{(1,1,\ldots,1)} = \kset~,\qquad (1,1,\ldots,1)_\nu = \nu~.
\end{equation}

With this notation at hand, equation \eqref{P:W3} becomes
\begin{equation}
	\Tr [A_{(k_1}A_{k_2}A_{k_3)}]=\frac12 \left[ 2 \scWL_{\vec{k}_{(3)}} -3 \scWL_{\vec{k}_{(2,1)}} + \scWL_{\vec{k}_{(1,1,1)}} \right]\,,
\end{equation}
and similarly for the other relations in \eqref{P:W1}--\eqref{P:W4}. In fact, we recognize the following pattern. With each term in the expansion of $\Tr [A_{(k_1}A_{k_2}\cdots A_{k_n)}]$ one can associate a unique partition $\nu$ of $n$. If $\nu$ has cycle number $c$, then the corresponding term involves the connected $c$-point correlator, which depends on the $c$-dimensional vector $\vec{k}_\nu$ defined in \eqref{C:k.mu.def}. The coefficient in front of this term is given by $(-1)^{c-1}|\mathcal{C}_\nu|/(n-1)!$, where $|\mathcal{C}_\nu|$ is the number of permutations of cycle type $\nu$.\footnote{Notice that the sum over all the coefficients for given $n$ vanishes.} Therefore, we find that equations \eqref{P:W1}--\eqref{P:W4} are generalized to 

%
\begin{equation}
\label{P:Wn}
	\Tr [A_{(k_1}A_{k_2} \cdots A_{k_n)}] = \sum\limits_{c=1}^n  \frac{(-1)^{c-1}}{(n-1)!} \sum\limits_{\nu \vdash n|c} 
		|\mathcal{C}_\nu| \scWL_{\kset_\nu}~. 
\end{equation}
This is our first result. In what follows, we will use it to express the generating functions $J_S$  and $J_A$ in terms of the connected Wilson loop correlators and prove Okuyama's observation \eqref{I:conjecture}.

Consider first $J_A$. We start from the exact expression \eqref{WL:J.MM} and expand the logarithm. This yields
\begin{equation}
\label{P:JA.1}
	N J_A \left( z, \lambda, \frac1N \right) = -\sum\limits_{n=1}^\infty \frac{(-z)^n}{n} \Tr (A_1^n)~.
\end{equation}
Using \eqref{P:Wn} with $\kset=(1,1,\ldots,1)$, taking account of \eqref{C:k.mu.spec} and reordering the summations over $n$ and $c$, \eqref{P:JA.1} becomes
\begin{equation}
\label{P:JA.W}
	N J_A \left( z, \lambda, \frac1N \right) = \sum\limits_{c=1}^\infty (-1)^c \sum\limits_{n=c}^\infty \frac{(-z)^n}{n!}\sum\limits_{\nu \vdash n|c} 
	|\mathcal{C}_\nu| \cWL_{\nu}~.
\end{equation}
This is our second result. Notice that the tilde for symmetrization has been dropped, because the symmetrization would have acted on the components of $\nu$, but the connected correlators are by definition symmetric in their arguments.

Next, consider $J_S$. Starting once more from \eqref{WL:J.MM} and expanding the logarithm gives
\begin{equation}
\label{P:JS.1}
	N J_S \left( z, \lambda, \frac1N \right) = -\sum\limits_{n=1}^\infty \frac{(-1)^n}{n} 
	\sum\limits_{k_1,k_2,\ldots,k_n=1}^\infty z^{k_1+k_2+\cdots +k_n}\Tr \left[A_{(k_1} A_{k_2}\cdots A_{k_n)}\right]~.
\end{equation}
Substituting \eqref{P:Wn} and reordering the sums one obtains
\begin{equation}
\label{P:JS.2}
	N J_S \left( z, \lambda, \frac1N \right) = \sum\limits_{c=1}^\infty \sum\limits_{n=c}^\infty \frac{(-1)^{n+c}}{n!}
	\sum\limits_{\nu \vdash n|c} |\mathcal{C}_\nu| \sum\limits_{k_1,k_2,\ldots,k_n=1}^\infty 
	z^{k_1+k_2+\cdots +k_n} \cWL_{\kset_\nu}~.
\end{equation}
Notice again that the tilde in the connected correlators can be omitted, this time, because the symmetrization is achieved implicitly by the summation over the $k$'s. 

The resulting sums can be simplified with the aid of the general identity
\begin{equation}
\label{P:sum.id}
	\sum\limits_{k_1,k_2,\ldots, k_s=1}^{\infty} f\left(\sum\limits_{i=1}^s k_i\right) 
	=\sum\limits_{k=1}^{\infty}\binom{k-1}{s-1}f(k)~,
\end{equation}
which needs to be applied $c$ times, once for each component of $\vec{k}_{\nu}$. Thus,
\begin{equation}
\label{P:sum.W}
	\sum\limits_{k_1,k_2,\ldots,k_n=1}^\infty z^{k_1+k_2+\cdots +k_n} \cWL_{\kset_\nu} 
	= \sum\limits_{k_1,k_2,\ldots,k_c=1}^\infty \left[ \prod\limits_{j=1}^c \binom{k_j-1}{\nu_j-1} \right] 
	z^{k_1+k_2+\cdots +k_c}\cWL_{\kset}~,
\end{equation}
where on the right hand side $\kset=(k_1,k_2,\ldots,k_c)$ is $c$-dimensional. We can now apply to \eqref{P:sum.W} the master formula \eqref{C:sum.f.master}, rendering
\begin{equation}
\label{P:sum.W1}
	\sum\limits_{k_1,k_2,\ldots,k_n=1}^\infty z^{k_1+k_2+\cdots +k_n} \cWL_{\kset_\nu} = 
	\sum\limits_{m=c}^\infty \frac{z^m}{m!} \sum\limits_{\mu \vdash m|c} |\mathcal{C}_\mu| \cWL_\mu 
	\sum\limits_{\sigma \in \SG_c} 
	\prod\limits_{j=1}^c \left[\binom{\mu_{\sigma(j)}}{\nu_j} \nu_j \right]~.
\end{equation}
The summation over the permutations $\sigma\in\SG_n$ explicits the necessary symmetrization. 
Let us remark that this expression is a function of the partition $\nu$. 

For the final steps of the calculation, we substitute \eqref{P:sum.W1} back into \eqref{P:JS.2} and exchange the summations over $m$ and $n$. This can be safely done because the binomials are non-zero only for $\mu_{\sigma(i)}\geq \nu_i$,  for all $i =1,2,\ldots,c$. Summing this over $i$ shows that only the terms with $m\geq n$ contribute to the sum. Hence,
\begin{equation}
\label{P:JS.3}
	N J_S \left( z, \lambda, \frac1N \right) = \sum\limits_{c=1}^\infty \sum\limits_{m=c}^\infty \frac{z^m}{m!}
	\sum\limits_{\mu \vdash m|c} |\mathcal{C}_\mu| \cWL_{\mu} 
	\sum\limits_{n=c}^\infty \sum\limits_{\nu \vdash n|c} \frac{(-1)^{n+c}}{n!} |\mathcal{C}_\nu|
	\sum\limits_{\sigma \in \SG_c} \prod\limits_{j=1}^c \left[\binom{\mu_{\sigma(j)}}{\nu_j} \nu_j \right]~.
\end{equation}
Here, starting from the sum over $n$, we recognize the right hand side of \eqref{C:sum.f.master} for the function\footnote{In \eqref{C:sum.f.master}, the symmetrization in $\tilde{f}$ is carried out over the $k$'s. In \eqref{P:JS.3}, because of the product over $j$, this is equivalent to symmetrizing the $\mu$'s.}
\begin{equation}
\label{P:f.master}
	f_{\kset} = (-1)^{k_1+k_2+\cdots +k_c +c}  
	\prod\limits_{j=1}^c \binom{\mu_j}{k_j}~.
\end{equation}
Therefore, we have
\begin{align}
\notag
	\sum\limits_{n=c}^\infty \sum\limits_{\nu \vdash n|c} \frac{(-1)^{n+c}}{n!} |\mathcal{C}_\nu|
	\sum\limits_{\sigma \in \SG_c} \prod\limits_{j=1}^c \left[\binom{\mu_{\sigma(j)}}{\nu_j} \nu_j \right]
	&= (-1)^c \sum\limits_{k_1,k_2,\ldots,k_c=1}^\infty (-1)^{k_1+k_2+\cdots +k_c}
	 \prod\limits_{j=1}^c \binom{\mu_j}{k_j}\\
\notag 
	&= \prod\limits_{j=1}^c\left[ - \sum\limits_{k_j=1}^\infty (-1)^{k_j} \binom{\mu_j}{k_j}\right]\\
	&= \prod\limits_{j=1}^c\left[ 1 - (1-1)^{\mu_j} \right] =1~.
\end{align}
Equation \eqref{P:JS.3} then reduces to 
\begin{equation}
\label{P:JS.W}
	N J_S \left( z, \lambda, \frac1N \right) = \sum\limits_{c=1}^\infty \sum\limits_{m=c}^\infty \frac{z^m}{m!}
	\sum\limits_{\mu \vdash m|c} |\mathcal{C}_\mu| \cWL_{\mu}~,
\end{equation}
which is our final result for $J_S$. Comparing \eqref{P:JS.W} with \eqref{P:JA.W} and taking into account \eqref{WL:W.series}, the relation \eqref{I:conjecture} between the generating functions follows immediately.

\section{Conclusions}\label{sec: concs}
In this paper, we revisited the generating functions of circular Wilson loops in $\mathcal{N}=4$ SYM in the totally symmetric and antisymmetric representations of the $U(N)$ gauge group, taking advantage of some previously known results derived from the Gaussian matrix model. The novelty of our work consists in using the connected correlators of multiply-wound Wilson loops in the fundamental representation, which were (re)introduced by Okuyama in \cite{Okuyama:2018aij}, as a basis.   
For the sake of concreteness, we report here the results for the generating functions,
\begin{align}
	J_S \left( z, \lambda, \frac1N \right)& = \frac{1}{N}\sum\limits_{c=1}^\infty \sum\limits_{n=c}^\infty \frac{z^n}{n!}
	\sum\limits_{\nu \vdash n|c} |\mathcal{C}_\nu| \cWL_{\nu}~,
	\\
	J_A \left( z, \lambda, \frac1N \right)& = \frac{1}{N}\sum\limits_{c=1}^\infty (-1)^c \sum\limits_{n=c}^\infty \frac{(-z)^n}{n!}\sum\limits_{\nu \vdash n|c} |\mathcal{C}_\nu| \cWL_{\nu}~.
\end{align}
In these equations, $\cWL_\nu$ denotes the connected correlators of multiply-wound Wilson loops, which are labelled by partitions $\nu$ (of $n$ into $c$ parts), and $|\mathcal{C}_\nu|$ is the number of permutations of the same cycle type as indicated by the partition $\nu$. When written in this fashion, the conjecture \eqref{I:conjecture} by Okuyama relating the two generating functions follows directly from the genus expansion \eqref{WL:W.series} for the connected correlators. 
We mention that our proof of the conjecture is exact in the 't Hooft coupling $\lambda$. As a by-product, we have computed the two-point connected correlator $\mathcal{W}_{(k_1,k_2)}$ to leading order in $1/N$ using the techniques introduced in \cite{CanazasGaray:2018cpk}, see appendix~\ref{app: W2}.

It is evident that the structure of the generating functions is governed by the representation theory of the symmetric group.\footnote{Recall that partitions label the conjugacy classes of the symmetric group.} In hindsight, this should not come as a surprise. The representation theory of the symmetric group is widely used in the analysis of matrix models. To be more precise, the symmetric polynomials appear as functions of $\e{g x_i}$, where $x_i$ are the eigenvalues that one integrates over. Thus, the (unconnected) correlators of multiply-wound Wilson loops translate to the power-sum symmetric polynomials, whereas a single Wilson loop in some representation $\mathcal{R}$ translates to the Schur polynomial correponding to that representation. Other bases of the symmetric functions could be equally used, because they are related by linear relations. For example, Fiol and Torrents \cite{Fiol:2013hna} used the monomial basis, from which the Schur basis is obtained by the Frobenius formula.

Our results suggest a number of further questions. We believe that \eqref{P:Wn} has a group theoretical origin, as it somewhat  resembles a relation between the elementary and power-sum symmetric polynomials. It should also be possible to generalize our results to arbitrary representations. A clear advantage of using the connected correlators as a basis, at least in the context of large $N$, is their genus expansion, which we indicated in \eqref{WL:W.series}. The $n$-point connected correlator has a leading term of order $N^{2-n}$, whereas all the traces of products of $A_k$ grow like $N$. Our calculation in appendix~\ref{app: W2} illustrates this point. If one starts from the exact solution for the matrix $A_k$, the first subleading term in its $\frac{1}{N}$-expansion is merely sufficient to calculate the leading term of the 2-point connected correlator, but nothing can be said for the 3- and higher-point connected correlators. Therefore, it would be very interesting to find an \emph{exact} solution of the matrix model that yields the connected correlators directly. By \emph{exact} solution we mean for finite $N$. This does not include the resolvent method, which crucially relies on a contiuum eigenvalue density in the large-$N$ limit. Another interesting question is the relation between the connected correlators and the $U(N)$ color invariants introduced in \cite{Fiol:2018yuc}. We leave these interesting lines of research for future work.

\appendix
\section{Combinatorial basics}\label{app: comb}
In this appendix, we review the minimal amount of concepts from combinatorial analysis that are necessary for our purposes. We also derive a ``master formula'' that converts sums over integers into sums over partitions. For a short overview, reference \cite{NIST} should suffice. A more detailed account can be found in \cite{Sagan_2001}.

The symmetric group $\SG_n$ consists of all bijections from the set $\{1,2,\ldots,n\}$ to itself using composition as group multiplication. The elements $\sigma\in \SG_n$ are called \emph{permutations}. The rank of $\SG_n$ is $n!$. 

Any permutation can be decomposed into a product of \emph{cycles}, which are nothing but cyclic permutations of disjoint subsets of $\{1,2,\ldots,n\}$. For example, in cycle notation, the permutation $\{1,2,3,4,5,6,7\}\rightarrow\{2,5,7,4,1,6,3\}$ would be written as $(251)(73)(4)(6)$. Depending on the permutation, the number of cycles can range from $1$ to $n$. If $\sigma$ contains $a_1$ fixed points (cycles of length one), $a_2$ cycles of length two, \ldots, and $a_n$ cycles of length $n$, then it is said to have \emph{cycle type}\footnote{In \cite{NIST}, the notation $(a_1, a_2, \ldots, a_n)$ is used to specify the cycle type, but this would be in conflict with the notation we introduce in the main body.} 
\begin{equation}
\label{C:cycle.type}
	t = \prod\limits_{\genfrac{}{}{0pt}{}{i=1}{(a_i>0)}}^{n} (i)^{a_i}~.
\end{equation}
The example above has $t=(3)^1(2)^1(1)^2$. Clearly, 
\begin{equation}
\label{C:n}
	n= \sum\limits_{i=1}^n i a_i~,
\end{equation}
and the number of cycles is given by 
\begin{equation}
\label{C:c}
	c = \sum\limits_{i=1}^n a_i~.
\end{equation} 
The cycle type characterizes a conjugacy class of $\SG_n$. 

A different way to indicate the cycle type, which turns out to be more useful for our purposes, is as a list $\nu= (\nu_1, \nu_2, \ldots \nu_c)$, with $\nu_i \in \{1,2,\ldots n\}$, of non-increasing integers, in which each integer $\nu_i$ is repeated $a_{\nu_i}$ times. In the example considered above, the permutation $(251)(73)(4)(6)$ is of cycle type $\nu=(3,2,1,1)\sim (3)^1(2)^1(1)^2=t$. It is evident that the list $\nu$ thus defined represents a \emph{partition} of $n$ into $c$ parts. We shall denote it by $\nu \vdash n|c$. Hence, there is a one-to-one correspondence between the cycle types of permutations $\sigma\in \SG_n$ and partitions of $n$. The number of permutations of cycle type $\nu \sim t$ (or the number of elements in the conjugacy class labelled by $\nu$) is given by
\begin{equation}
\label{C:C.mu}
	|\mathcal{C}_\nu|  = \frac{n!}{\prod\limits_i i^{a_i} a_i!}~.
\end{equation} 
 
Now, in this paper we encounter sums of the form 
\begin{equation}
\label{C:sum.f}
	\sum\limits_{k_1,k_2,\ldots, k_c=1}^{\infty} f_{\kset}~,
\end{equation}
where $\kset=(k_1,k_2,\ldots,k_c)$ is a $c$-dimensional vector of integers and $f_{\kset}$ a function of $\kset$. Defining the symmetrized function
\begin{equation}
\label{C:f.sym}
	\tilde{f}_{\kset} = \frac1{c!} \sum\limits_{\sigma\in \SG_c} f_{\sigma(\kset)}~,
\end{equation}
with $\sigma(\kset)_i = k_{\sigma(i)}$, and assuming that the summations can be carried out in any order, the above sum is clearly equal to
\begin{equation}
\label{C:sum.f.sym}
	\sum\limits_{k_1,k_2,\ldots, k_c=1}^{\infty} f_{\kset} = \sum\limits_{k_1,k_2,\ldots, k_c=1}^{\infty} \tilde{f}_{\kset}~.
\end{equation}
Because $\tilde{f}$ is explicitly symmetric in its arguments, we can order them by size in a non-increasing manner. This way, $\kset=(k_1,k_2,\ldots,k_c)$ becomes a partition of $n=k_1+k_2+\cdots+k_c$ into $c$ parts, $\nu\vdash n|c$. Collecting identical terms in the sum and noting that their multiplicity is given by the multinomial coefficient
\begin{equation}
\label{C:mult.coeff}
	\binom{c}{a_1,a_2,\ldots, a_c}= \frac{c!}{a_1! a_2! \cdots a_c!}~,
\end{equation}
where the integers $a_i$ are determined by the cycle type corresponding to $\nu$, yields
\begin{equation}
\label{C:sum.f.sym2}
	\sum\limits_{k_1,k_2,\ldots, k_c=1}^{\infty} f_{\kset} = \sum\limits_{n=c}^\infty \sum\limits_{\nu \vdash n|c} 
	\frac{c!}{\prod\limits_i a_i!} \tilde{f}_{\nu}~.
\end{equation}
Furthermore, we can express the multinomial coefficients \eqref{C:mult.coeff} in terms of \eqref{C:C.mu} and rewrite the result using 
the identity 
\begin{equation}
\label{C:prod.ident}
	\prod\limits_{i=1}^n i^{a_i} = \prod\limits_{i=1}^c \nu_i \qquad \text{for } \nu \vdash n|c~. 
\end{equation}
Thus, we obtain the master formula
\begin{equation}
\label{C:sum.f.master}
	\sum\limits_{k_1,k_2,\ldots, k_c=1}^{\infty} f_{\kset} = \sum\limits_{n=c}^\infty \frac{c!}{n!}
		\sum\limits_{\nu \vdash n|c} |\mathcal{C}_\nu| \left(\prod\limits_{i=1}^c \nu_i \right) \tilde{f}_\nu~.
\end{equation}
\section{Connected 2-point correlator}\label{app: W2}
As a by-product of our investigation, we have calculated the connected 2-point correlator $\cWL_{(k_1,k_2)}$ using the techniques and results of \cite{CanazasGaray:2018cpk}. Because the calculation is quite short, we would like to present it here. 

The starting point is the matrix $A(k)$ defined in \eqref{WL:A:expl}. In the nomenclature of \cite{CanazasGaray:2018cpk}, this corresponds to the matrix elements in the \emph{number basis}. It is convenient to change to the \emph{position basis} \cite{CanazasGaray:2018cpk}, in which the leading term in $1/N$ is diagonal,
\begin{equation}
\label{W2:A.pos}
	A(k)_{ij} = \e{k\sqrt{\lambda}\sqrt{1+\frac1{2N}} \cos\theta_i} \delta_{ij} + \frac1N \sum\limits_{r,s=1}^\infty
	\frac{\sin(r\theta_i)}{\sin\theta_i} \frac{\sin(s\theta_j)}{\sin\theta_j} \BesselI[r+s](k\sqrt{\lambda}) + 
	\Order\left(N^{-2}\right)~. 
\end{equation}  
Here, $\theta_i$ are given implicitly by the equation
\begin{equation}
	\left(N+\frac{1}{2}\right)\left(\theta_i-\sin\theta_i\cos\theta_i\right)=\left(i-\frac{1}{4}\right)\pi,
	\qquad
	i=1,2,\ldots,N\,,
\end{equation}
and $\BesselI[n](z)$ denotes the modified Bessel functions of the first kind. In the large-$N$ limit, sums over $i$ can be converted into integrals via
\begin{equation}
	\frac{1}{N}\sum_{i=1}^Nf(\theta_i)=\frac{2}{\pi}\left(1+\frac{1}{2N}\right)\int_0^{\pi}\textrm{d}\theta\sin^2\theta f(\theta)-\frac{1}{4N}\left[f(0)+f(\pi)\right]+\Order\left(N^{-2}\right)\,.
\end{equation}
This conversion is necessary to compute products of matrices as well as traces.

As a first check, by a calculation similar to (C.6) in \cite{CanazasGaray:2018cpk}, we obtain the 1-point function
\begin{equation}
\label{W2:A.tr}
	\cWL_{(k)} = \Tr A(k) = \frac{2N}{k\sqrt{\lambda}} \BesselI[1](k\sqrt{\lambda}) + \Order\left(N^{-1}\right)~.
\end{equation}
Notice that there is no term of order $N^0$, in agreement with \eqref{WL:W.series}. For the connected 2-point correlator we introduce the matrix
%
%
\begin{equation}
\label{W2:W.def}
	W(k_1,k_2) = A(k_1+k_2) - A(k_1)A(k_2)~.
\end{equation}
In order to calculate its matrix elements in the position basis, some properties of the modified Bessel functions are needed, in particular, the generating function and summation theorem
\begin{equation}
\label{W2:I.gen and sum}
	\e{x\cos \theta} = \sum\limits_{k=-\infty}^\infty \BesselI[k](x) \cos(k\theta)\,,
	\qquad
	\BesselI[n](x+y) = \sum\limits_{k=-\infty}^\infty \BesselI[n-k](x) \BesselI[k](y)~.
\end{equation}
One obtains
\begin{equation}
\label{W2.W.pos}
	W(k_1,k_2)_{ij} =  \frac1N \sum\limits_{r,s=1}^\infty
	\frac{\sin(r\theta_i)}{\sin\theta_i} \frac{\sin(s\theta_j)}{\sin\theta_j}
	\sum\limits_{t=0}^{\infty} \BesselI[r+t](k_1\sqrt{\lambda})\BesselI[s+t](k_2\sqrt{\lambda}) + \Order\left(N^{-2}\right)~. 
\end{equation}
Taking the trace gives the connected 2-point correlator 
\begin{align}
\notag
	\cWL_{(k_1,k_2)} &=  
	\sum\limits_{r=1}^{\infty} \sum\limits_{t=0}^{\infty} \BesselI[r+t](k_1\sqrt{\lambda})\BesselI[r+t](k_2\sqrt{\lambda}) \\
\notag
	&= \sum\limits_{v=1}^{\infty} v \BesselI[v](k_1\sqrt{\lambda})\BesselI[v](k_2\sqrt{\lambda}) \\
\notag
	&= \frac{k_1 k_2}{k_1+k_2} \left( \frac1{k_1} + \frac1{k_2} \right) \sum\limits_{v=1}^{\infty} 
		v \BesselI[v](k_1\sqrt{\lambda})\BesselI[v](k_2\sqrt{\lambda}) \\
\notag
	&= \frac12 \sqrt{\lambda} \frac{k_1 k_2}{k_1+k_2} \sum\limits_{v=1}^{\infty} \left\{ 
	\left[ \BesselI[v-1](k_1\sqrt{\lambda}) -\BesselI[v+1](k_1\sqrt{\lambda}) \right] \BesselI[v](k_2\sqrt{\lambda}) \right.\\
\notag & \quad \qquad \left.
	+ \BesselI[v](k_1\sqrt{\lambda})\left[ \BesselI[v-1](k_2\sqrt{\lambda}) -\BesselI[v+1](k_2\sqrt{\lambda}) \right]
	\right\} \\
	&=  \frac12 \sqrt{\lambda} \frac{k_1 k_2}{k_1+k_2} \left[ \BesselI[0](k_1\sqrt{\lambda}) \BesselI[1](k_2\sqrt{\lambda}) +
	\BesselI[1](k_1\sqrt{\lambda}) \BesselI[0](k_2\sqrt{\lambda}) \right]~.
\label{oku:Z.tr}
\end{align}
This reproduces equation (3.10) of \cite{Okuyama:2018aij}.

We remark that it is not possible to calculate any higher-point correlators using this approach; one would not only need the $1/N^2$ terms in \eqref{W2:A.pos}, but also carefully take into account further $1/N$ corrections that stem from matrix multiplications, as explained in \cite{CanazasGaray:2018cpk}.

\bibliography{NotesWL}
\end{document}